\documentclass[aps,prb,reprint,showpacs,superscriptaddress]{revtex4-2}

\usepackage{graphicx}
\usepackage{bm}
\usepackage[utf8]{inputenc}
\usepackage{amsmath}
\usepackage{float}
\usepackage{amsmath} 
\usepackage{booktabs}
\usepackage[x11names,table,xcdraw]{xcolor}
\usepackage{comment}

\usepackage{soul}

\begin{document}
\preprint{APS/123-QED}

\title{Domain-wall Magnetic-texture dependent Creep Motion \\driven by Spin-transfer Torques}

\author{Lucas Javier Albornoz}
\affiliation{Laboratoire de Physique des Solides, Universit\'e Paris-Saclay, CNRS, 91405 Orsay, France.}
\affiliation{Instituto Balseiro, Univ. Nac. Cuyo - CNEA, Av. Bustillo 9500 (R8402AGP), S. C. de Bariloche, Río Negro, Argentina.}
\affiliation{Instituto de Nanociencia y Nanotecnología (CNEA-CONICET), Nodo Bariloche, Av. Bustillo 9500 (R8402AGP), S. C. de Bariloche, Río Negro, Argentina}
\affiliation{Gerencia de Física, Centro Atómico Bariloche, Av. Bustillo 9500 (R8402AGP), S. C. de Bariloche, Río Negro, Argentina.}

\author{Rebeca D\'iaz Pardo}
\affiliation{Laboratoire de Physique des Solides, Universit\'e Paris-Saclay, CNRS, 91405 Orsay, France.}

\author{Aristide Lemaître}
\affiliation{Centre de Nanosciences et de Nanotechnologies (C2N), CNRS, Universit\'e Paris-Saclay, 91120 Palaiseau, France.}

\author{Sebastian Bustingorry}
\affiliation{Instituto de Nanociencia y Nanotecnología (CNEA-CONICET), Nodo Bariloche, Av. Bustillo 9500 (R8402AGP), S. C. de Bariloche, Río Negro, Argentina}
\affiliation{Gerencia de Física, Centro Atómico Bariloche, Av. Bustillo 9500 (R8402AGP), S. C. de Bariloche, Río Negro, Argentina.}

\author{Javier Curiale}

\affiliation{Instituto de Nanociencia y Nanotecnología (CNEA-CONICET), Nodo Bariloche, Av. Bustillo 9500 (R8402AGP), S. C. de Bariloche, Río Negro, Argentina}
\affiliation{Gerencia de Física, Centro Atómico Bariloche, Av. Bustillo 9500 (R8402AGP), S. C. de Bariloche, Río Negro, Argentina.}
\affiliation{Instituto Balseiro, Univ. Nac. Cuyo - CNEA, Av. Bustillo 9500, 8400 S. C. de Bariloche, Rio Negro, Argentina.}

\author{Vincent Jeudy}
\affiliation{Laboratoire de Physique des Solides, Universit\'e Paris-Saclay, CNRS, 91405 Orsay, France.}

\date{\today}

\begin{abstract}
We explore the contributions of adiabatic and non-adiabatic spin-transfer torques (STT) of a spin-polarized current to the thermally activated creep motion of domain-walls in a thin (Ga,Mn)(As,P) film with perpendicular anisotropy.
For a domain-wall transverse to current, the non-adiabatic STT is found to act as an external magnetic field. Close to the compensation between these two terms, the adiabatic contribution is strongly enhanced. The domain-wall velocity may be both increased or reduced by the adiabatic STT, which we associate to variations of creep pinning energy barrier with domain-wall magnetic texture. 
Far from compensation, the contribution of adiabatic STT is negligible. Field and current driven domain-wall motion present common universal behaviors described by the quenched Edwards Wilkinson universality class.  
\end{abstract}

\pacs{Valid PACS appear here}
\maketitle


\section{Introduction}

The dynamics of magnetic domain-walls (DWs) driven by spin polarized electrical current have been the subject of a large number of fundamental and applied researches in the past decades due to expected applications to spintronic devices~\cite{parkin_naturenano_2015_race_track,grolier_nat_elec_2020}. When a current is crossing a DW, the spin transfer from charge carriers to the local magnetic moments of atoms may be phenomenologically described by adiabatic and non-adiabatic contributions~\cite{Zhang_prl_2004,thiaville_EPL_2005}. The dynamics of DW results from the coupling between its velocity and internal magnetization precession frequency. In thin films with perpendicular anisotropy, for a DW transverse to current, the non-adiabatic torque (na-STT) is equivalent to a perpendicular magnetic field, and controls the velocity, while the precession of DW internal magnetization depends both on adiabatic spin transfer torque (a-STT) and na-STT. A so-called Walker threshold is expected to separate DW motion with a steady magnetic texture at low drive from that with a precessional texture at high drive~\cite{curiale_prl_2012_spin_drift}. In the presence of disorder inherent to magnetic materials, the pinning strongly modifies DW dynamics. Despite a fairly abundant theoretical~\cite{duine_prb_2008,ryu_prb_2011,jin_pre_2021}, and experimental~\cite{Yamanouchi_science_2007,adam_prb_2009,lee_PRL_2011,moon_prl_2013,duttagupta_natphys_2015,diaz_pardo_prb_2019} literature on this topic, the respective contributions of a-STT and na-STT to DW depinning and dynamical behaviors remain an open issue.

The competition between DW interaction with random disorder, its elasticity, the driving force $f$ (associated to field or current), and thermal noise leads to rich universal behaviors~\cite{chauve_2000,lemerle_prl_1998,diaz_pardo_prb_2019}, shared by a large variety of interfaces moving in random media~\cite{agoritsas_physicaB_2012}.
Below a depinning threshold, DWs move in the thermally activated creep regime. The velocity scales as $\ln v \sim f^{-\mu}$, where $\mu$ is the universal creep exponent. Field driven DW motion in thin ferro- and ferri-magnets~\cite{savero-torres_prb_2019,albornoz_prb_2021} is well described by the value $\mu=1/4$~\cite{lemerle_prl_1998,jeudy_prl_2016}. This value coincides with prediction for a one-dimension elastic line, with short range elasticity and disorder interactions, moving in a two-dimensional medium~\cite{chauve_2000} described by the so-called quenched Edwards Wilkinson universality class~\cite{edwards_wilkinson_1982}. 
Current driven DW motion presents slightly different universal behaviors. In extended geometry, the DWs are observed to form mountain-like structures~\cite{moon_prl_2013,diaz_pardo_prb_2019}, which reflects the directionality of current-DW interaction causing a decrease of the drive-magnitude with increasing DW tilting~\cite{diaz_pardo_prb_2019}. However, for DWs kept transverse to current, magnetic field and current driven DW motion were found to present common creep~\cite{lee_PRL_2011,diaz_pardo_prb_2019} and depinning universal behaviors~\cite{diaz_pardo_prb_2019}. Note that in tracks, the tilting of DWs~\cite{Yamanouchi_science_2007,duttagupta_natphys_2015}, Joule heating~\cite{Boulle_PRL_2008,San_Emeterio_prl_2010,lee_PRL_2011,haltz_prb_2019_Joule} and the contribution of edge pinning as revealed by dome-like DW-shapes~\cite{duttagupta_natphys_2015,herrera-diez_prb_2018} may lead to effective creep exponents different from the universal value $\mu=1/4$. 

In this context, the respective contributions of the STTs to creep motion remain controversial. Duine and Morais Smith~\cite{duine_prb_2008} and later Ryu {\it et al}~\cite{ryu_prb_2011} suggested that a key ingredient in addition to DW elasticity could be its magnetic texture. In their model, developed for a DW transverse to current,
the pinning potential depends both on the DW position and its internal magnetization tilting angle. They found that the na-STT is equivalent to a magnetic field as for free DW motion~\cite{thiaville_EPL_2005}. 
Surprisingly, for the contribution of a-STT to DW magnetization tilt, both Refs.~\cite{duine_prb_2008,ryu_prb_2011} consider only spatial fluctuations of DW anisotropy due to disorder while the Bloch/Néel DW anisotropy, which favors a magnetization direction parallel to DW is not taken into account. Moreover, the contribution of a-STT is assumed to enhance DW pinning~\cite{duine_prb_2008,ryu_prb_2011}, while it could be at the origin of non-trivial creep dynamical behaviors~\cite{lecomte_PRB_2009_internal_degree_freedom}.  
On the experimental side, the equivalence between na-STT and magnetic field seems rather well verified~\cite{adam_prb_2009,lee_PRL_2011}. In contrast, determining the contribution of a-STT to DW creep motion remains an experimental challenge due to Joule heating which may vary significantly the sample temperature~\cite{San_Emeterio_prl_2010,lee_PRL_2011,haltz_prb_2019_Joule}, and impedes stringent tests of theoretical predictions.


Here, we report a study on DW creep motion in an extended geometry, and for which Joule effects are precisely avoided. 
We enlighten a compensation between magnetic field and na-STT over one order of magnitude of current density, which is compatible with a frozen direction of DW-magnetization. Close to the compensation, the contribution of a-STT is found to increase and also reduce DW velocity, which rules out a contribution of a-STT only dominated by DW pinning. The model of energy barrier that we develop to account for experimental observations relies on combined contributions of magnetic field, na-STT, and a-STT to DW motion and magnetic texture.
Section II presents the experimental methods and a phenomenological description of DW motion. In section III, we discuss the zero-crossing of DW velocity and analyze the contribution of non-adiabatic STT. Section IV is dedicated to the role played by adiabatic STT on the DW magnetic texture and creep motion. 

\section{Phenomenology of Domain-Wall motion}
\subsection{Experimental Methods}
\begin{figure*}[ht!]
	\begin{center}
		\includegraphics[width=1.9\columnwidth]{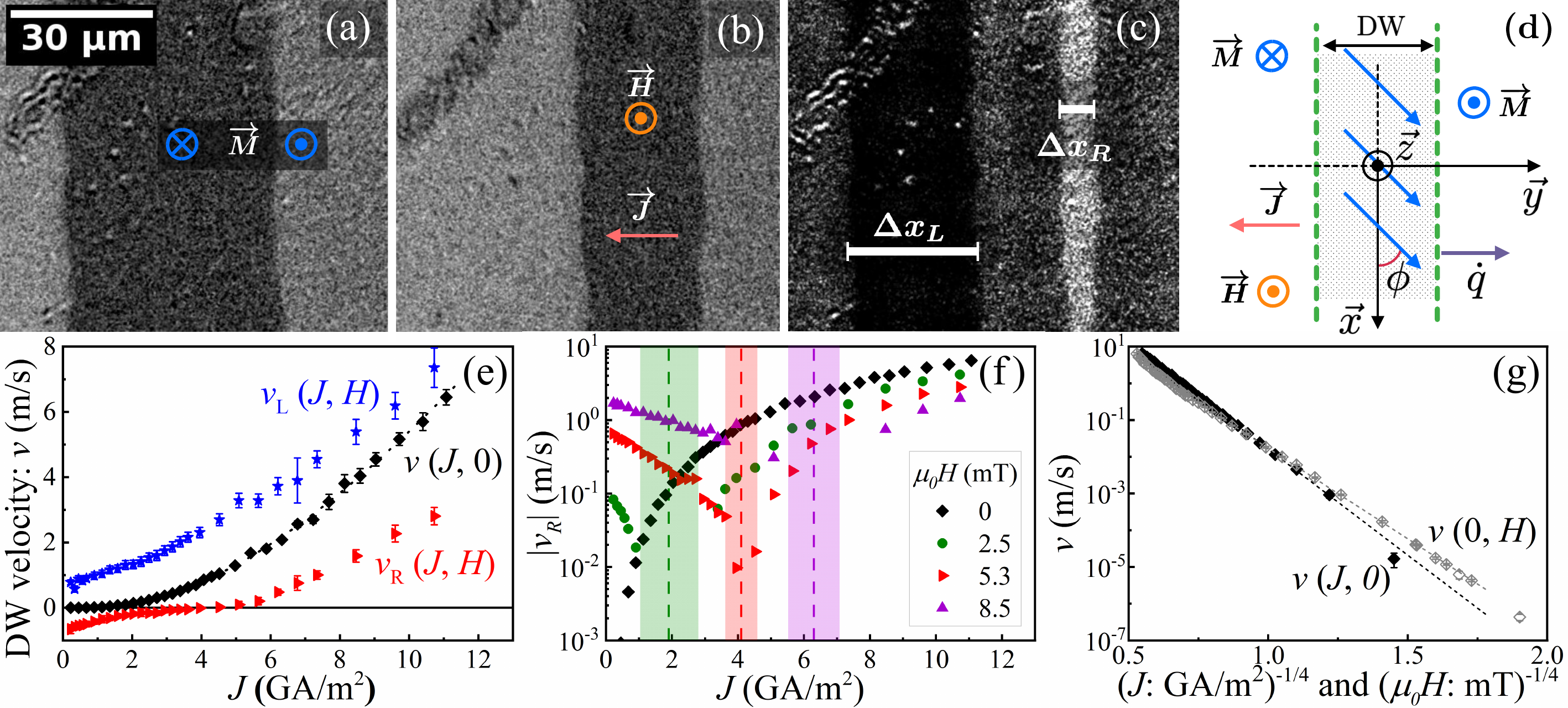} 
		\caption{Phenomenology of Domain-Wall Dynamics. 
		(a-c) Displacement of magnetic domain walls produced by simultaneous magnetic-field and current pulses, and observed by MOKE microscopy. In (a) and (b), the two gray levels correspond to opposite magnetization directions $\vec{M}$, perpendicular to the film. The direction of current density  $\vec{J} = - J \hat{y}$ (with $J > 0$) and magnetic field  $\mu_0\vec{H} = \mu_0 H \hat{z}$ (with $H > 0$) are indicated by arrows in (b). (a) Initial state with three domains separated by two DWs perpendicular to $\vec{J}$. (b) Same DWs after 10 simultaneous current and field pulses of amplitudes $ J = 4.5\,\mathrm{GA/m^2} $ and $ \mu_0 H = 2.1\,\mathrm{mT} $ and duration $ \Delta t = 1.4\,\mathrm{\mu s} $. (c) Subtraction of image (a) from (b) enlightening the different mean displacements of the left ($ \Delta x_L $) and right ($ \Delta x_R $) DWs.
		(d) Diagram of a DW magnetic texture, showing in particular the magnetization tilting angle $\phi$ at the DW center.
		(e) Velocities versus current density $J$ of the left ($v_L(J,H)$) and right ($v_R(J,H)$) DWs for a finite ($\mu_0 H = 5.3\,\mathrm{mT}$) and a zero ($v(J,0)=v_L(J,0)=v_R(J,0)$) applied magnetic field.  
		(f) Absolute value of $v_R(J,H)$ versus $J$ for different values of $\mu_0 H$. The semi-log plot emphasizes the drop down of velocity. The vertical dashed lines indicate the values of current density $J_c$ at the compensation condition where the velocity vanishes and the colored shaded surface areas correspond to the associated uncertainties.
		(g) Independent field ($v(0,H)$) and current ($v(J,0)$) driven DW velocity in semi-log scale versus $(\mu_0 H)^{-1/4}$, and $J^{-1/4}$, respectively, evidencing a common scaling creep behavior. All the experiments are performed at $T=55 \,\mathrm{K}$.}   
		\label{fig:pmoke}
	\end{center}
\end{figure*}
The DW motion was studied in a 4~nm thick film of a diluted magnetic semiconductor forming part of a stack made of AlAs(30)/GaAs(2)/(Ga,Mn)(As,P)(3)/(Ga,Mn)As(1) (the numbers in parentheses are the thicknesses in nm), grown by molecular beam epitaxy on a (001) GaAs/AlAs buffer ~\cite{niazi_apl_2013}. The film has a perpendicular anisotropy and a Curie temperature $T_C = 65\,\mathrm{K}$. It was patterned by electron lithography into a rectangle shape of size 133 $\times$ 210$\mu$m$^2$. Two 40 $\mu$m wide gold electrodes separated by 110$\mu$m, were deposited by evaporation parallel to the narrow sides of rectangle. They were used to generate an homogeneous current density producing DW motion by STT (see~\cite{diaz_pardo_prb_2019} for details).
The current pulse amplitude varied between 0 and 11~$\,\mathrm{GA/m^2}$.
We verified that the Joule effect had a negligible contribution on DW dynamics~\cite{diaz_pardo_prb_2019,curiale_prl_2012_spin_drift}. Perpendicular magnetic field pulses of adjustable amplitude (0-65~mT) were produced by a $\approx$ 75 turns small coil (diameter $\sim$ 1 mm) mounted on the sample.
The sample was fixed in an optical He-flow cryostat allowing a temperature regulation between 5.7~K and $T_C$. To observe the dynamics of the DWs, we used a magneto-optical Kerr effect microscope in polar configuration (resolution $\sim$ 1~$\mu m$). The mean displacement of DWs ($\Delta x$) was produced by magnetic field and/or current pulses of duration ($\Delta t$), which could be varied between 1~$\mu$s and 120~s. The DW velocities are defined as the ratio $v=\Delta x/\Delta t$. The phenomenology of domain-wall motion is described in Fig.~\ref{fig:pmoke}, for a temperature $T=55 \,\mathrm{K}$.

\subsection{Universal Creep Dynamics}

Independent magnetic field and current driven DW motion is compared in Fig.~\ref{fig:pmoke}(g). As can be observed, both velocity curves, $v(H, J=0)$ and $v(H=0,J)$, present a good agreement with the creep scaling law $\ln v \sim H^{-\mu}$, with the same value for the creep exponent ($\mu=1/4$). Therefore the DW dynamics are compatible with common universal behavior, as discussed in details in Ref.~\cite{diaz_pardo_prb_2019}.
Note that the two curves (for which $ \mu_0 H^{-\mu}$ and  $J^{-\mu}$ share the same scale) cross so that a single homothety between $ \mu_0 H$ and $J$ is not sufficient to superimpose the whole velocity curves, as assumed in previous works~\cite{adam_prb_2009,lee_PRL_2011}. The observed slightly different slopes originate from the different material dependent pinning parameters controlling field and current driven DW motion~\cite{diaz_pardo_prb_2019}.

\subsection{Combined Current and Field Driven Dynamics}

The motion of DWs produced by combined current and magnetic field pulses (of the same duration) are reported in Figs.~\ref{fig:pmoke}(a-c). The initial state (see Fig.\ref{fig:pmoke}(a)) consists of two parallel DWs (called left and right DW in the following) separating a central domain with magnetization $\vec{M}$ perpendicular to the film and two domains with opposite magnetization directions. A combined current and magnetic field pulse pushes differently the two DWs (see Fig.\ref{fig:pmoke}(b)).
The current moves the two DWs in the same direction, opposite to the current flow (towards the right side), as expected for (Ga,Mn)As~\cite{curiale_prl_2012_spin_drift,adam_prb_2009}, while the magnetic field changes the distance between the two DWs (the central domain width is reduced due to the opposite direction between magnetic field $\vec{H}$ and magnetization $\vec{M}$).
As a result (see Fig.\ref{fig:pmoke}(c)), the DW mean displacement of the left ($\Delta x_L$) (right ($\Delta x_R$)) DW is larger (smaller) due to the additive (opposite) contributions of current and field.

\subsection{Zero crossing of Domain-Wall Velocity} 

More insights into dynamics are shown in Fig.\ref{fig:pmoke} (e), which compares the left and right DW velocities ($v_{L,R}=\Delta x_{L,R}/\Delta t$) versus current density ($J$). As expected, the additive (subtractive) contributions of the field and current  enhance (reduce) the velocity of left ($v_L(J,H)$) (right ($v_R(J,H)$)) DW compared to the zero field velocity $v(J, 0)$. Interestingly, as the current density is reduced, the velocity $ v_R $ is observed to go to zero (for $J\approx 4\,\mathrm{GA/m^2}$) and to change of sign. Highlights of this phenomenon are reported in Fig.\ref{fig:pmoke}(f), which shows absolute value of $v_R$ in semi-log scale versus $J$ for different fixed applied magnetic fields. As it can be observed, the value of current density ($J_c$) at which the compensation occurs systematically increases with increasing applied magnetic field value. Notice also that the velocity curves present an asymmetry with respect to $J_c$, which enlightens the role played by adiabatic STT on the DW magnetic texture and creep motion, as discussed later.

\section{Non-Adiabatic Spin-Transfer Torque and Domain-Wall Dynamics}

\subsection{Compensation between Field and Current}

In order to discuss the compensation between magnetic field and current, we have measured systematically the values of $J_c$ corresponding to the zero crossing of DW velocity ($\lvert v_R \rvert \rightarrow 0$) for a set of fixed applied magnetic field value $\mu_0H$, and different temperatures close to $T_C$.
\begin{figure}[th!]
	\begin{center}
		\includegraphics[width=1.0\columnwidth]{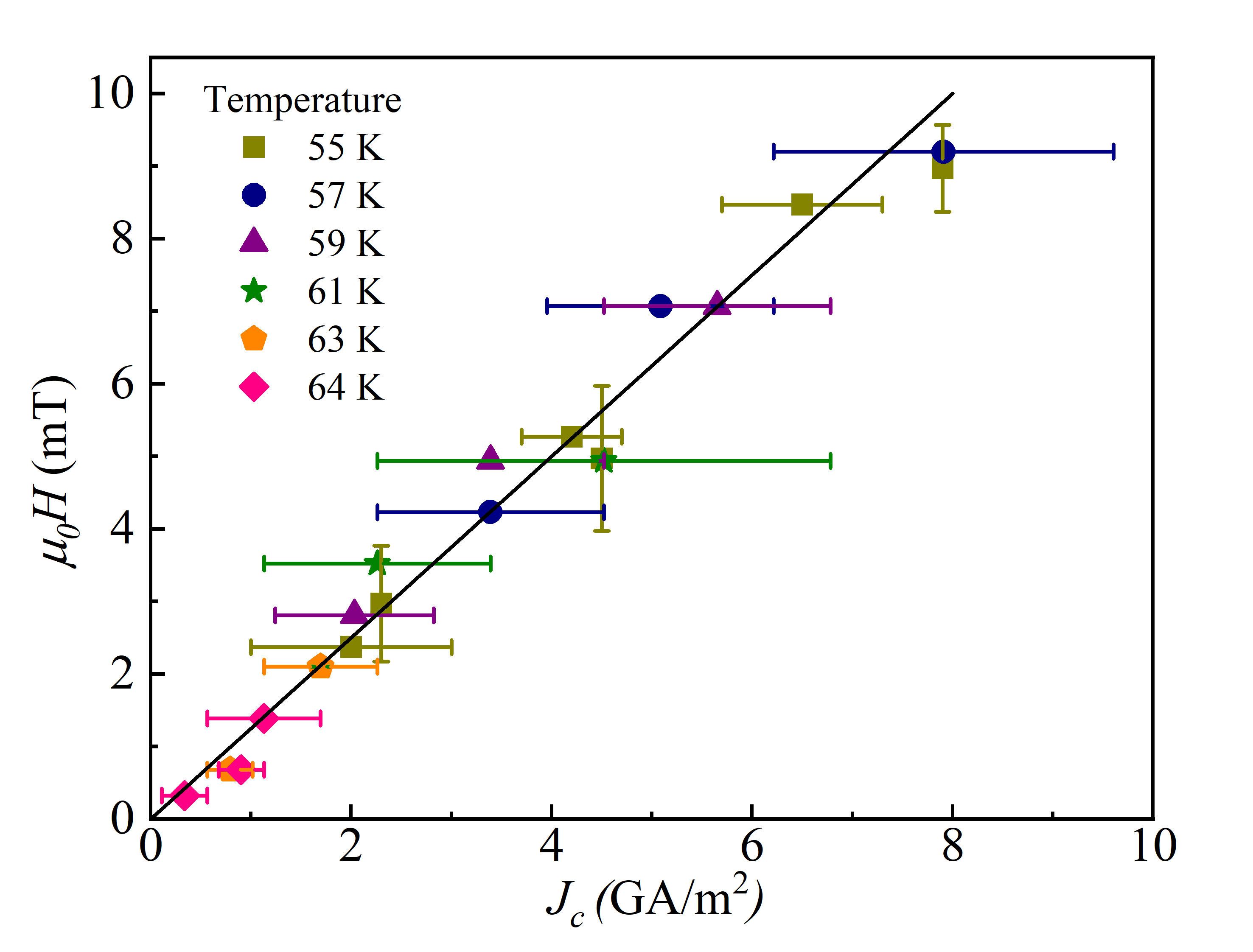}
		\caption{Current density $J_c$ versus applied magnetic field at the compensation condition corresponding to the zero crossing of DW velocity ($\lvert v_R \rvert \rightarrow 0$). The data obtained for different temperatures close to $T_C$ collapse on a linear variation, whose slope is $\mu_0 H/J_c=1.3\pm 0.2~\mathrm{mT/(GA/m^2)}$.
		}
		\label{fig:Compensation}
	\end{center}
\end{figure}
The results are reported in Fig.~\ref{fig:Compensation} for the temperature range $T/T_C>0.8$. As it can be observed, the data cover an order of magnitude in magnetic field values ($0.8 <\mu_0 H<9~\mathrm{mT}$) and collapse onto a single master curve. Over this field range, $\mu_0 H$ is found to be essentially proportional to  $J_c$, with a slope $\mu_0 H/J_c=1.3 \pm 0.2~\mathrm{mT/(GA/m^2)}$.    
%
Moreover, as discussed next, the single slope also strongly suggests that the DW texture remains in the same magnetic state over the whole explored magnetic field range. 
%
%
Note that our analysis of the zero crossing of DW velocity gives direct access to the equivalent torques exerted by magnetic field and current, without any assumption on analogies and differences between the velocity curves proposed in previous works~\cite{Yamanouchi_science_2007,duttagupta_natphys_2015,San_Emeterio_prl_2010,lee_PRL_2011,haltz_prb_2019_Joule}. 

\subsection{Study of the compensation}

For the analysis of DW motion, we consider a film of thickness $t$ with infinite lengths in the $\vec{x}$ and $\vec{y}$ direction, and a perpendicular magnetic anisotropy $K_0$. An infinite DW (almost parallel to the $x$-axis, see Fig.~\ref{fig:pmoke}(d)) separates two domains of magnetization $\vec{M}$ with opposite directions parallel ($+M_s$), and anti-parallel ($-M_s$) to $\vec{z}$, respectively. As for the right DW (see Figs.\ref{fig:pmoke}(a-b)), a magnetic field aligned in the direction $\vec{z}$ tends to push the DW along $-y$. A spin polarized current of density $\vec{J} = -J \hat{y}$ (with $J > 0$) is flowing in the direction $-y$ and tends to push the DW along $+y$. The DW position is defined by $q(x)$, and the tilting of magnetization by the angle $\phi(x)$, with $\phi(x)=0$ corresponding to a Bloch DW.
Following Refs.~\cite{duine_prb_2008,ryu_prb_2011}, the effective potential per unit surface area determining the force exerted on the DW can be written: $\Sigma=\Sigma_q+\Sigma_{\phi}+\Sigma_{dis}$, where $\Sigma_{dis}$ is the contribution of the random pinning disorder. The two other terms are:
\begin{gather}
\Sigma_q=\frac{\sigma}{2}\left(\frac{d q}{d x}\right)^2+2\mu_0M_s(H-\beta\chi J)q
\label{Sigm_q},\\
\Sigma_{\phi}=2\Delta A \left(\frac{d \phi}{d x}\right)^2+2K\Delta \sin^2\phi-2\mu_0 M_s \Delta \chi J\phi.
\label{Sigm_phi}
\end{gather}
In Eqs.~\eqref{Sigm_q} and \eqref{Sigm_phi}, $\sigma$ is the DW surface energy, and $\Delta$ its  thickness parameter, $A$ the exchange energy constant, and  $\beta$ the non-adiabatic spin transfer parameter. $K$ ($=\mu_0M_s H_K/2$) is the (Bloch-Néel) anisotropy energy of the DW, with $H_K=N_x M_s$ the anisotropy field, and $N_x$ ($\approx t/(t+\Delta)$) is the demagnetizing factor of the DW~\cite{mougin_epl_2007}. The product $\gamma \Delta \chi J$ is the spin drift velocity~\cite{thiaville_EPL_2005} with
\begin{equation}
\chi =\dfrac{g\mu_BP}{2 \vert e \vert \Delta M_s \gamma},
	\label{eq:chi}
\end{equation}
where $g$ is the Landé factor, $\mu_B$ the Bohr magneton, $e$ the electron charge, and $P$ the spin polarization of current. 
In Eq.~\eqref{Sigm_q}, the first term describes the elasticity of the DW, and the second term the contribution of the magnetic field and na-STT to the DW drive. The three terms in Eq.~\eqref{Sigm_phi} reflect the contributions of exchange interaction along the DW, DW Bloch-Néel anisotropy, and a-STT, respectively.
Notice here that left and right DWs are related by symmetry and thus are Eqs.~\eqref{Sigm_q} and \eqref{Sigm_phi}: the Zemman and the a-STT terms for the left DW reads $-2\mu_0M_s(H+\beta\chi J)q$ and $+2\mu_0 M_s \Delta \chi J\phi$, respectively. In the following we focus on the right DW using Eqs.~\eqref{Sigm_q} and \eqref{Sigm_phi}.

For the analysis of compensation between magnetic field and current (see Fig.~\ref{fig:Compensation}), we start with the so-called q-$\phi$ model~\cite{thiaville_EPL_2005}, which describes the free motion of a DW ($\Sigma_{dis}=0$). The DW is assumed to be straight ($dq/d x=0$) and its magnetization tilt to be homogeneous ($d\phi/d x=0$). The motion is governed by the Slonczewski Eqs.~\cite{slonczewski_aip_proc_1972,thiaville_EPL_2005}:
\begin{equation}
	\begin{array}{lr}
	-\alpha\frac{\dot{q}}{\Delta} +\dot{\phi}=\frac{\gamma}{2 \mu_0 M_s}\frac{\delta \Sigma}{\delta q}\\
	-\frac{\dot{q}}{\Delta} -	\alpha\dot{\phi}=\frac{\gamma}{2 \mu_0 M_s \Delta}\frac{\delta \Sigma}{\delta \phi},
	\end{array}
\label{eq:Slonczewski}
\end{equation}
where $\gamma$ is the gyromagnetic ratio, coupling the DW velocity $\dot{q}$ and magnetization precession frequency $\dot{\phi}$~\footnote{The proper equations for the left DW are obtained by symmetry using $\dot{q} \to -\dot{q}$, $J \to -J$ and $\phi \to \phi + \pi$.}.
Inserting Eqs.~\eqref{Sigm_q} and \eqref{Sigm_phi} into Eqs.~\eqref{eq:Slonczewski}, and assuming a motion in the steady state ($\dot{\phi}=0$) leads to the well known result:
\begin{gather}
	\dot{q}=-\frac{\gamma \Delta}{\alpha}(H-\beta\chi J)
	\label{qpointST},\\
	\alpha \frac{H_K}{2} \sin (2 \phi)=H-\beta\chi J+ \alpha \chi J,
	\label{phiST}
\end{gather}
%
which allows to examine qualitatively the contribution of a-STT and na-STT to DW motion. At the compensation condition the velocity $\dot{q} \rightarrow 0$ (see Eq.~\eqref{qpointST}), the magnetic field and na-STT compensate ($H-\beta\chi J=0$). The proportionality factor $\mu_0 H/J = \mu_0\beta \chi$, where $\chi$ is given by Eq.~\eqref{eq:chi}, is expected to be constant provided the magnetic state of the DW remains steady.
%
Moreover, for $\dot{q}> 0$, $H-\beta\chi J$ and $\alpha \chi J$ are positive. Both exert torques increasing the magnetization tilt angle $\lvert \phi \rvert$ (see Eq.~\ref{phiST}) and consequently the DW Bloch-Néel anisotropy energy.
On the contrary, for $\dot{q}< 0$, the torques exerted by $H-\beta\chi J$ and $\alpha \chi J$ have opposite signs and thus the a-STT tends to maintain a small value of $\lvert \phi \rvert$.

\subsection{DW Magnetic State at the Compensation}

Let us now compare the ratio $\mu_0 H/J=1.3 \pm 0.2 ~\mathrm{mT/(GA/m^2)}$ deduced from Fig.~\ref{fig:Compensation} to the prediction at compensation for a steady DW magnetic texture ($\dot{\phi}=0$). For the ferromagnetic semiconductor (Ga,Mn)As, close to
$T_C$ ($0.8<T/T_C<1$), the ratio between spin polarization and magnetization is athermal: $P/M_s \approx 0.027   \mathrm{(kA.m)^{-1}}$~\cite{curiale_prl_2012_spin_drift}. $\Delta=2.5 \pm 1.0 ~\mathrm{nm}$~\cite{dourlat_PRB_2008} and $\beta \approx 0.25$~\cite{curiale_prl_2012_spin_drift} do not vary with temperature.
Therefore, the ratio $(\mu_0 H/J)_{st}= \beta \chi=\beta g \mu_B P/(\gamma \Delta 2\vert e \vert M_s)$ is expected to be athermal as observed in Fig.~\ref{fig:Compensation}. The predicted value  $(\mu_0 H/J)_{st}=0.5-1~\mathrm{mT/(GA/m^2)}$ ($g=2$, $\gamma =1.76 \times 10^{11}~\mathrm{Hz/T}$) is in rather good agreement with experimental results, as already reported in Ref.~\cite{adam_prb_2009}.
In contrast, for the asymptotic precessional state (i.e., a time average of $\cos(2 \phi)$ equal to $0$ in Eq.~\ref{Sigm_phi}), the predicted ratio $(\mu_0 H/J)_{prec}= (\mu_0 H/J)_{st}(1+1/(\beta \alpha))=9-15~\mathrm{mT/(GA/m^2)}$ is significantly larger than the measurement. This suggests that at the compensation between magnetic field and na-STT, the DW presents a (fluctuating) magnetic texture remaining close to the steady state. 

Note that the ratio $\mu_0 H/J=0.2-8 \times 10^{-2}~\mathrm{ mT/(GA/m^2)}$ reported in the literature~\cite{Boulle_PRL_2008,San_Emeterio_prl_2010,lee_PRL_2011} for Pt/Co/Pt films or multilayers  is between one and two orders of magnitude lower than our measurement ($1.3 \pm 0.2 ~\mathrm{mT/(GA/m^2)}$). Comparing the material dependent parameters occurring in the product $\beta \chi$ (=$\mu_0 H/J$) (see Eq.~\eqref{eq:chi} for $\chi$) for Pt/Co/Pt~\cite{Boulle_PRL_2008,San_Emeterio_prl_2010,lee_PRL_2011}  ($\Delta=6-9~\mathrm{nm}$, $P\approx 0.5$, and $\beta \approx 0.35-1.5$, $M_s~\approx 1.4~\mathrm{MA/m}$), and for (Ga,Mn)As ($P\approx 0.3$, and $M_s~\approx 10~\mathrm{kA/m}$ for $T/T_C=0.9$~\cite{curiale_prl_2012_spin_drift}), we see that the dominant difference is the magnetization which is two orders of magnitude larger for Pt/Co/Pt than for (Ga,Mn)As.
Therefore, the much larger STT efficiency in (Ga,Mn)As may be essentially attributed to the scaling of $\mu_0 H/J=\mu_0\beta \chi$ with $1/M_s$.
%

Moreover, the range ($0.8 <\mu_0 H<9~\mathrm{mT}$) over which the linear variation is observed in Fig.~\ref{fig:Compensation} has to be compared to the Walker limit separating the steady and precessional state. Indeed, since $\lvert \sin 2 \phi \rvert \leq 1$ in Eq.~\eqref{phiST}, the steady state is expected to occur only over limited ranges of magnetic field and current. At the compensation between field and na-STT ($H-\beta\chi J=0$), the Walker limit reads $\mu_0 H_w=\mu_0 \beta\chi J_w = \alpha \mu_0 H_K/2 \approx 2~\mathrm{mT}$ (for $T/T_C=0.9$). Without pinning, the slope $\mu_0 H/J$ should change by about one order of magnitude ($\approx (\mu_0 H/J)_{prec}/(\mu_0 H/J)_{st}$) below and above $\mu_0 H_w$. Therefore, the observation (in Fig.~\ref{fig:Compensation}) of a constant ratio $\mu_0 H/J$ over the whole range of explored magnetic field indicates that the DW remains in the steady DW state above the Walker limit, which strongly suggests that the pinning disorder impedes the precession of DW magnetization.

\section{Adiabatic Spin-Transfer Torque and Domain-Wall Dynamics}

\subsection{Effective driving magnetic field}
\begin{figure}[th!]
	\begin{center}
		\includegraphics[width=0.95\columnwidth]{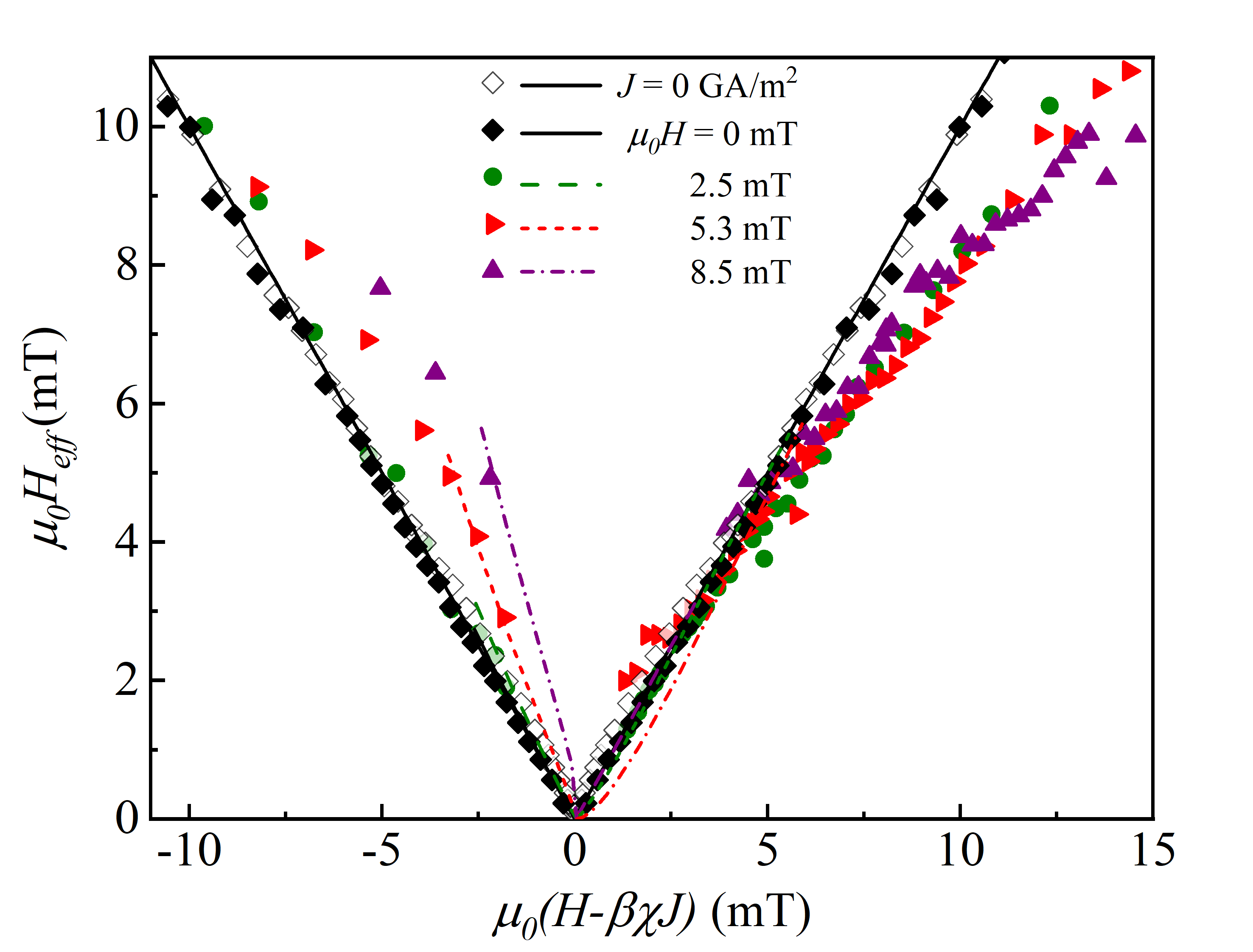}
		\caption{Contribution of adiabatic-STT (a-STT) to DW creep motion. Effective magnetic field $H_{eff}$ versus $\mu_0(H-\beta \chi J)$, with $\beta \chi=1.3 ~\mathrm{mT/(GA/m^2)}$ for  $J=0~\mathrm{GA/m^2}$ (empty diamonds) and different applied magnetic field values: $\mu_0 H=0~\mathrm{mT}$(filed gray diamonds), $2.5~\mathrm{mT}$ (olive circles), $5.3~\mathrm{mT}$ (red right triangles), and $8.5~\mathrm{mT}$ (purple up  triangles). The data points are deduced from velocity curves (see Fig.~\ref{fig:pmoke} (e-g)) and Eq.~\eqref{eq:Heff}. For $J=0~\mathrm{GA/m^2}$ and $\mu_0 H=0~\mathrm{mT}$, the superposition of curves is compatible with $\mu_0H_{eff}=|\mu_0(H-\beta \chi J)|$, which reveals a negligible contribution of a-STT. Increasing the applied magnetic field enlightens a-STT effects: $\mu_0H_{eff}$ is enhanced (reduced)  for $\mu_0(H-\beta \chi J) <0$ ($>0$). The different lines are fits of Eq.~\eqref{eq:effectiveH2}, using the DW energy of Eq.~\eqref{eq:sigmaSTsmall} and are obtained for a single adjustable parameter $p$ (see text).}
		\label{fig:Heff}
	\end{center}
\end{figure}

We now discuss the contribution of a-STT to the creep motion from the velocity curves shown in Figs. Fig.~\ref{fig:pmoke}(e-f)). In the creep regime, the DW velocity is described by~\cite{jeudy_prl_2016}:
\begin{equation}
v(H_{eff})=v(H_d)\exp \left(-\frac{\Delta E}{k_B T}\right),
\label{eq:creep}
\end{equation}
where $\Delta E=k_B T_d((H_{eff}/H_d)^{-\mu}-1)$ is the effective pinning energy barrier, $\mu=1/4$ the universal creep exponent, and $k_B T$ the thermal activation energy. In Eq.~\eqref{eq:creep}, $k_B T_d$, $H_d$ and $v(H_d)$ are material dependent parameters characterizing the height of effective pinning barrier, the depinning threshold and velocity, respectively. In the following (see Fig.~\ref{fig:Heff}), the effective field $H_{eff}(H,J)$ is assumed to describe both the effect of magnetic field and current on the motion and is used to compare experimental results and theoretical predictions.
From Eq.~\eqref{eq:creep}, the effective magnetic field can written as a function of the DW velocity $v(H,J)$:
\begin{equation}
	H_{eff}(H,J)=H_d \left[1-\frac{T}{T_d} \ln \large\vert \frac{v(H,J)}{v(H_d)}\large\vert \right]^{-4}.
	\label{eq:Heff}
\end{equation}
In order to define uniquely the variations of $H_{eff}(H,J)$ from the velocity curves $v(H,J)$, a set of material dependent parameters was fixed from a fit with Eq.~\eqref{eq:creep} of the velocity curve $v(H,0)$ (see Fig.~\ref{fig:Compensation}(f)) obtained for field driven DW motion ($J=0$). Assuming $v(H_d)=5.5~\mathrm{m/s}$~\cite{diaz_pardo_prb_2019},
the other pinning parameters are $T_d=400~\mathrm{K}$, $H_d=13~\mathrm{mT}$.
For field driven DW motion (with $J=0$), the effective field is simply $H_{eff}(H,0)=\vert H \vert$ and corresponds to two straight lines in Fig.~\ref{fig:Heff}.

For current driven DW ($H=0$), the effective field $H_{eff}(0,J)$ is plotted as a function of $-\mu_0 \beta \chi J$, where we used the value of $\beta \chi$ ($= 1.3~\mathrm{mT/(GA/m^2)}$) deduced from the compensation between field and na-STT (see section III). As it can be observed in Fig.~\ref{fig:Heff}, the two curves $H_{eff}(H,0)$ and $H_{eff}(0,J)$ are almost perfectly superimposed. This is the expected behavior since the velocity curves $v(H,0)$ and $v(0,J)$ shown in Fig.~\ref{fig:pmoke}(f) are close to overlap. 
The superposition of curves considering only the na-STT indicates that the contribution of a-STT is negligible for current driven DW dynamics with $H=0$. This is compatible with results reported in Ref.~\cite{diaz_pardo_prb_2019}, which show that for a DW perpendicular to the current, current and magnetic field driven DW motion follow close universal behaviors for the creep and depinning regimes, corresponding to the quenched Edwards Wilkinson universality class.

For the combined field and current driven DW motion, $\mu_0 H_{eff}(H,J)$ was deduced from the velocity curves obtained for the right and left DW (see Fig.~\ref{fig:pmoke}(d-e)). The results are plotted in Fig.~\ref{fig:Heff} as a function of $\mu_0(H-\beta \chi J)$ for the right DW and $\mu_0(-H-\beta \chi J)$ for the left DW, for different fixed magnetic field values of $\mu_0 H$.
The curves present striking asymmetric features: for the right DW $\mu_0 H_{eff}$ is increased (reduced) for negative (positive) values of $\mu_0(H-\beta \chi J)$ (i.e., of DW velocity).
Close to the compensation $\mu_0(H-\beta \chi J) \approx 0 mT$, raising $\mu_0 H$ increases the necessary current density ($J \approx H/(\beta \chi))$ to achieve the compensation condition ($v \to 0$) and consequently enhances the contribution of a-STT ($\alpha \chi J$, see Eq.~\eqref{phiST}). Therefore, the asymmetric behavior observed in Fig.~\ref{fig:Heff} reflects the contribution to DW creep motion of the a-STT, which may both increase or decrease the effective field (i.e. the effective pinning energy barrier, see Eq.~\eqref{eq:creep}). This observation is in contradiction with the assumption~\cite{duine_prb_2008, ryu_prb_2011} of an a-STT contribution only increasing the pinning energy barrier.

As discussed qualitatively in subsection III. B, assuming steady motion without disorder, for $\dot{q}> 0$ one has that $H-\beta\chi J$ and $\alpha \chi J$ are positive and tend to increase the DW Bloch-Néel anisotropy energy, while for $\dot{q}< 0$, $H-\beta\chi J$ and $\alpha \chi J$ have opposite sign and might change the sign of the energy constribution.
With disorder, in the creep regime, increasing the DW energy increases the pinning barrier heights and reduces the DW velocity. Therefore, the a-STT is expected to reduce DW velocity. On the contrary, for $\dot{q}< 0$, the a-STT tends to maintain a small value of $\lvert \phi \rvert$ and consequently to enhance DW velocity.
Though this argument is based on the non-disordered case, it qualitatively indicates that the contribution of a-STT to DW dynamics should be asymmetric on both sides of the compensation with a reduced (enhanced) velocity for $\dot{q}>0$ ($\dot{q}<0$).

\subsection{Model for the Creep Motion}

In order to analyze more quantitatively the contribution of a-STT and DW texture to creep motion, we use standard scaling arguments~\cite{lemerle_prl_1998,agoritsas_PRE_13}. The free energy per unit thickness of a DW segment of length $L$, deformed over a distance $u$~\cite{lemerle_prl_1998,hartmann_prb_2019} can be written:
\begin{equation}
	\delta F= \frac{\sigma}{2} \frac{u^2}{L} -2 \mu_0 M_s \vert H -\beta \chi J \vert u L -\delta F_{dis} \pm \Sigma_{\phi DW} L,
	\label{eq:freeenergy}
\end{equation}
where the first term is the elastic energy produced by the increase of DW length. The second term is the gain of Zeeman energy due to magnetization reversal, the absolute value ensures a reduction of the free energy regardless of the sign of $H -\beta \chi J$ (i.e. of the direction of motion). The third term $\delta F_{dis}$ is the contribution of DW pinning. In addition to those terms commonly used to describe the depinning of an elastic line~\cite{lemerle_prl_1998,hartmann_prb_2019}, we introduce a term $\Sigma_{\phi DW} L$ associated to the variation of DW magnetic texture with the current. The prefactor $\pm$ ($ = sgn(H-\beta \chi J )$ of this term accounts for its dependency on the relative direction of current and DW motion (as discussed in the previous section).
 
Let us now discuss the magnetic texture contribution $\Sigma_{\phi DW}$. Equation~\eqref{Sigm_phi} describes the variation of magnetization angle along the DW, and in particular the structure of Bloch lines~\cite{buttiker_pra_1981,malozemoff} and their displacement with the current. A theoretical discussion of this effect on the creep motion is beyond the scope of this paper. In the following, we restrict ourselves to the simplest DW texture consisting in a uniform ($d \phi/d x=0$) and steady  magnetization direction ($\dot{\phi} = 0$). 
For a moving DW whose magnetic texture is at equilibrium one can assume that $\delta \Sigma_{\phi}/\delta \phi=0$. Then Eq.~\eqref{Sigm_phi} leads to:
\begin{equation}
\frac{H_K}{2}\sin 2\phi = \chi J,
\label{eq:eqphi}
\end{equation} 
%
This equation indicates that the a-STT essentially changes the DW magnetization direction. In comparison to Eq.~\eqref{phiST}, the terms $ H -\beta \chi J$ has disappeared since we are considering an equilibrium solution instead of the dynamics as described by the Slonczewski equations. This is effectively equivalent to assuming that there is no coupling between the DW position $q$ and magnetization angle $\phi$. (The Euler Eqs. $\delta \Sigma_q/\delta q=0$, and  $\delta \Sigma_{\phi}/\delta {\phi}=0$ were solved separately.)
%
Inserting Eq.~\eqref{eq:eqphi} into Eq.~\eqref{Sigm_phi} leads to the energy at equilibrium per unit thickness and length of a DW:
\begin{equation}
	\Sigma_{\phi_{st}}= -\frac{1}{2}\mu_0 H_K M_s \Delta \left[\sqrt{1-h^2}+h \arcsin (h) -1 \right], 
	\label{eq:sigmaST}
\end{equation}
where $h=2\chi J/H_K$.  
Interestingly, the limit of small tilting angle ($2\phi \approx h$, see Eq.~\eqref{eq:eqphi}):
\begin{equation}
	\Sigma_{\phi_{st}}(\phi \rightarrow 0) = -\frac{\Delta \mu_0}{N_x} \left(\chi J \right)^2
	\label{eq:sigmaSTsmall}
\end{equation}
remains close to Eqs.~\eqref{eq:sigmaST} ($\Sigma_{\phi_{st}}/\Sigma_{\phi_{st}}(\phi \rightarrow 0)\leq \pi-2$) up to the Walker limit ($\phi=\pm \pi/4$), and therefore enlightens the close-scaling of DW anisotropy energy with $J^2$.


We can now derive the effective magnetic field. With the usual assumption of a DW displacement following the power-law variation $u=u_0 (L/L_c)^\zeta$, where $\zeta=2/3$ is the roughness exponent of the DW~\cite{lemerle_prl_1998,diaz_pardo_prb_2019}, and $L_c$ the so-called collective pinning length, Eq.~\eqref{eq:freeenergy} can be written:
\begin{equation}
	\delta F(L)= AL^{1/3}-B\vert H-\beta \chi J\vert L^{5/3} \pm \Sigma_{\phi_{st}} L, 
	\label{eq:freeenergy2}
\end{equation}
with $A=\sigma u_0^2/(2L_c^{4/3})$, and  $B=2 \mu_0 M_s u_0/L_c^{2/3}$. 
The height of the effective energy barrier is deduced from $\partial \delta F(L=L_{opt})/ \partial L=0$, which yields:
%
\begin{equation}
	L_{opt}=\left(\frac{A}{5B}\right)^{3/4} \left(\mp p \Sigma_{\phi_{st}} +\sqrt{(p \Sigma_{\phi_{st}})^2+ \vert H-\beta \chi J\vert }\right)^{-3/2}.
	\label{eq:Lopt}
\end{equation}

In Eq.~\eqref{eq:Lopt}, we have introduced the constant  $p=3/\sqrt{20AB}$, which depends on the micromagnetic and pinning parameters of the material.
Inserting $L_{opt}$ into Eq.~\eqref{eq:freeenergy2} leads to the energy barrier
\begin{equation}
\Delta E (H_{eff})=\frac{4}{5} \left(\frac{A^5}{5BH_{eff}}\right)^{1/4},
\label{eq:barrier}
\end{equation}
with the effective field
\begin{equation}
	H_{eff}=\vert H-\beta \chi J \vert \frac{\left[\mp r + \sqrt{1+r^2}\right]^{10}}{\left[1+\frac{5}{3}r \left(r \mp \sqrt{1+r^2 } \right) \right]^{4}}.
	\label{eq:effectiveH2}
\end{equation}
The ratio  
\begin{equation}
 	r=\frac{p \Sigma_{\phi_{st}}(J)}{\sqrt{\vert H-\beta \chi J\vert }},
 	\label{eq:ratio}
\end{equation}
present in Eq.~\eqref{eq:effectiveH2}, suggests an enhanced contribution of a-STT and DW magnetic texture to the creep motion close to compensation ($H-\beta \chi J \rightarrow 0$) between magnetic field and na-STT. Far from the compensation, current driven DW motion is expected to be compatible with the universal behavior of magnetic field driven DW ($\mu=1/4$), as observed in Fig.~\ref{fig:pmoke}, and already reported in Ref.\onlinecite{diaz_pardo_prb_2019}. Eqs.~\eqref{eq:barrier}, \eqref{eq:effectiveH2}, and ~\eqref{eq:ratio} predict a deviation form the qEW universal behavior for sufficiently large values of the ratio $r$.

\subsection{Adiabatic-STT and DW magnetic texture}

A quantitative comparison between the predictions for effective field $H_{eff}$ and experimental data is reported in Fig.~\ref{fig:Heff}. For the set of curves obtained for different fixed magnetic fields $H$, we performed a global fit of Eq.~\eqref{eq:effectiveH2}, including the variation of ratio $r$ (Eq.~\eqref{eq:ratio}) and $\Sigma_{\phi_{st}}(\phi \rightarrow 0)$ (Eq.~\eqref{eq:sigmaSTsmall}). Note that since the value of $\beta \chi$ ($= 1.3~\mathrm{mT/(GA/m^2)}$) is fixed, the global fit relies on a single adjustable parameter $p=(2.0 \pm 0.3) \times 10^6~\mathrm{T^{-1} (A.m)^{-1/2}}$.
As it can be observed, the prediction presents a rather good agreement with the data close to the compensation (see the dashed lines in Fig.~\ref{fig:Heff}). This result demonstrates (see Eq.~\eqref{eq:eqphi}) that the adiabatic-STT essentially controls the tilt of DW magnetization (and more generally, the steady DW magnetic texture).
%


As written previously, the Walker limit would be $\mu_0 H_w \approx 2~\mathrm{mT}$ without pinning. The fact that data remain in agreement with prediction for the steady DW state for a larger value ($\mu_0 H_{eff} \approx 4-5~\mathrm{mT}$) suggests a contribution of the random pinning potential 
to maintain a fixed direction of magnetization, as already argued for the compensation.

It is interesting to assess the variation with micromagnetic and pinning parameters of the a-STT contribution to creep motion from standard scaling arguments. The collective pinning length $L_c$ at depinning $H=H_d$ can be deduced from Eq.~\eqref{eq:freeenergy2}, assuming $AL_c^{1/3} \sim B H L_c^{5/3}$. Replacing $A$ and $B$ by their expressions, the scaling relation $u_0 \sim \Delta$, yields $L_c \sim (\sigma \Delta/(4\mu_0 M_s H_d))^{1/2}$ and 	$p \sim 3/(4\Delta \mu_0 M_s\sqrt{5H_d})$. Moreover, at the compensation ($H-\beta \chi J=0$), the contribution of DW magnetic anisotropy should scale with the DW elasticity  $\Sigma_{\phi_{st}} L_c \sim AL_c^{1/3}$  (see Eq.~\eqref{eq:freeenergy2}), which leads to $\mu_0 H \sim \beta \mu_0 \sqrt{M_s H_d N_x}$
Therefore, the contribution of a-STT is expected to be larger, and to occur at lower field for material presenting a low saturation magnetization and weak depinning field.



\section{Conclusion}
Our experimental and theoretical work clarifies the contributions of magnetic field, non adiabatic-STT, and adiabatic-STT to the creep motion of domain wall and highlights the interplay between domain wall magnetic texture and adiabatic-STT.

For the effective pinning barrier height, the magnetic field and na-STT play a similar role. The a-STT introduces a new contribution proportional to the square of current density, which is associated to domain wall anisotropy. Its sign is positive (negative) when the a-STT contributes to increase (decrease) the domain-wall anisotropy energy and its magnitude is large only close to the compensation between magnetic field and na-STT.
 
Moreover, the creep motion is found to be closely related to the state of the magnetic texture.  
Close to the compensation, the magnetic texture correspond to a quasi steady state with magnetization tilt angle fluctuating around a mean angle controlled by the a-STT. The data are compatible with the steady state well above the Walker limit predicted for a free DW, which strongly suggests that the random pinning potential depends on DW magnetic texture. 
Above the Walker limit, the magnetic texture is expected to present more complex behaviors with probably nucleation and motion of Bloch lines, which should be particularly interesting to study. 
  
Implications for the creep motion of interfaces presenting universal behaviors are important since we show that the internal degree of freedom of an elastic interface (the magnetic texture of DW) produces deviation form the prediction qEW universality class.

\acknowledgements
We wish to thank A. Thiaville for careful reading of the manuscript. V. J., J. C., and S. B. acknowledge support by the France-Argentina project ECOS-Sud No. A19N01. 

\bibliography{refs_creep_DMI}

\end{document}